\pdfoutput=1
\documentclass[aps,pre, twocolumn, groupedaddress]{revtex4-1}

\usepackage{lipsum}
\usepackage{mathtools}
\usepackage{graphicx}
\usepackage{dcolumn}
\usepackage{amsmath}    
\usepackage{amssymb}
\usepackage{bm}
\usepackage{hyperref}
\usepackage{latexsym}
\usepackage{verbatim}
\usepackage{color}

\usepackage[caption=false]{subfig}

\def\beq{\begin{equation}}
\def\eeq{\end{equation}}

\def\beq{\begin{equation}}                          
\def\eeq{\end{equation}}                          
\def\bea{\begin{eqnarray}}                          
\def\eea{\end{eqnarray}}

\DeclareRobustCommand{\uvec}[1]{{%
  \ifcsname uvec#1\endcsname
     \csname uvec#1\endcsname
   \else
    \bm{\hat{\mathbf{#1}}}%
   \fi
}}

\preprint{}

\begin{document}

\title{Active nematics with quenched disorder}

\author{Sameer Kumar}
\email[]{sameerk.rs.phy16@itbhu.ac.in}
\affiliation{Indian Institute of Technology (BHU), Varanasi, U.P. India - 221005}

\author{Shradha Mishra}
\email[]{smishra.phy@itbhu.ac.in}
\affiliation{Indian Institute of Technology (BHU), Varanasi, U.P. India - 221005}

\begin{abstract}
We introduce a two-dimensional active nematic with quenched disorder. We write the
coarse-grained hydrodynamic equations of motion for slow variables,
{\em viz.} density and orientation.  Disorder strength is tuned from zero to large values.
Results  from the numerical solution of equations of motion as well as  the calculation of two-point orientation correlation 
function using linear approximation,  
shows that the  ordered steady-state  follows a disorder dependent crossover from  
quasi long-range order (QLRO)  to short-range order (SRO).
Such crossover is due to the pinning of $\pm 1/2$ topological defects in the presence of finite disorder, which 
breaks the system in uncorrelated domains.
Finite disorder slows the dynamics of $+1/2$ defect, and it
 leads to slower growth dynamics. 
The two-point correlation functions for the density and orientation fields show good dynamic scaling but no static scaling for the different disorder strengths.
Our findings can motivate experimentalists to verify the results and find applications in living and artificial apolar systems in the presence of a quenched disorder.

\end{abstract}

\maketitle
{\em Introduction}: 
Dynamics and steady-state of a collection of active self-propelled particles 
with different kinds of inhomogeneities has become an interesting area of research 
\cite{naturecommindrani, naturecommmorin, shradhapre2018, jtoner2018prl, jtoner2018pre, fparuniprl2013, AnanyoMaitra2019}. 
Recent studies have mostly focused on the polar 
self-propelled particles in the presence of inhomogeneous agents/medium \cite{dombrowski2004, sanchez2012, sumino2012}.
The effect of disorder in active polar particles introduces many exciting features, 
which, in general, do not present in the corresponding equilibrium
system of the same symmetry \cite{imryma1975}. Studies on the effect of disorder in apolar particles are limited to the equilibrium system only \cite{Rotunno2005}.
Disorders are present almost everywhere inactive apolar systems \cite{naturecommindrani}, but 
ordering and steady-state of active apolar particles with the disorder is rarely studied. \\

Variety of systems where  particles have head-tail symmetry, 
like vibrated granular rods  \cite{kudrolli2003, tsimiring2007}, collection of molecular motors, 
cytoskeletal filaments  \cite{melanocytes, bsubtilis}, mesenchymal, 
epithelial cells monolayers \cite{Duclos2017, Kawaguchi2017, Saw2017, BlanchMercader2018},  
bacterial colonies \cite{Doostmohammadi2016, DellArciprete2018, Yaman2019}, 
and colonies of swarming filamentous bacteria \cite{Li2019} are a few examples of the active apolar system. 
The collection of such active apolar particles,  forming an orientationally ordered state, is called $active \  nematics$. 
Most of the previous active nematic studies are
for a clean system  \cite{shradhanjop, husechateprl2004, sriramadititonerepl2000, Doostmohammadi2018}.  But, inhomogeneity or
disorder can play a crucial role in steady-state and kinetics of active nematics, which is our current study's focus.\\

In this letter, we study quenched disorder's effect on a collection of active apolar particles on a 
two-dimensional substrate. The disorder introduced as a random field of strength $h_0$ in the coarse-grained 
hydrodynamic equations of motion for slow variables; local density $\rho({\bf r}, t)$  and  order parameter  $\mathcal{Q}({\bf r}, t)$.
We first characterize the steady-state and then study the ordering kinetics.
We calculate the nematic order parameter (NOP)  ${\bf Q}$ {\em vs.} system 
size $N$ for different  $h_0$. 
For clean or homogeneous active nematic, NOP decay algebraically with $N$ (quasi-long range order, QLRO). 
But for a finite disorder, NOP shows a power-law  decay, for small $N$ and a disorder dependent crossover to an exponential decay (short-range order, SRO) 
for large $N$, and the same we confirm  by the  calculation of
two-point orientation correlation function in the steady-state, 
using a linear approximation. 
The origin of such crossover for the finite disorder ($h_0 \ne 0$) is due to the pinning of the $\pm 1/2$ defects.  For large enough $N$, it breaks the system in uncorrelated domains, and the size of these domains depends on the disorder strength. 
Although the orientation field is significantly affected due to  disorder, the density fluctuation 
remains unaffected and shows the usual giant number fluctuation (GNF) \cite{shradhaprl, husechateprl2004, sriramadititonerepl2000} 
for all disorder strengths ($h_0$).\\

\begin{figure*}      
\mbox{\includegraphics[width=8 cm,height=3cm]{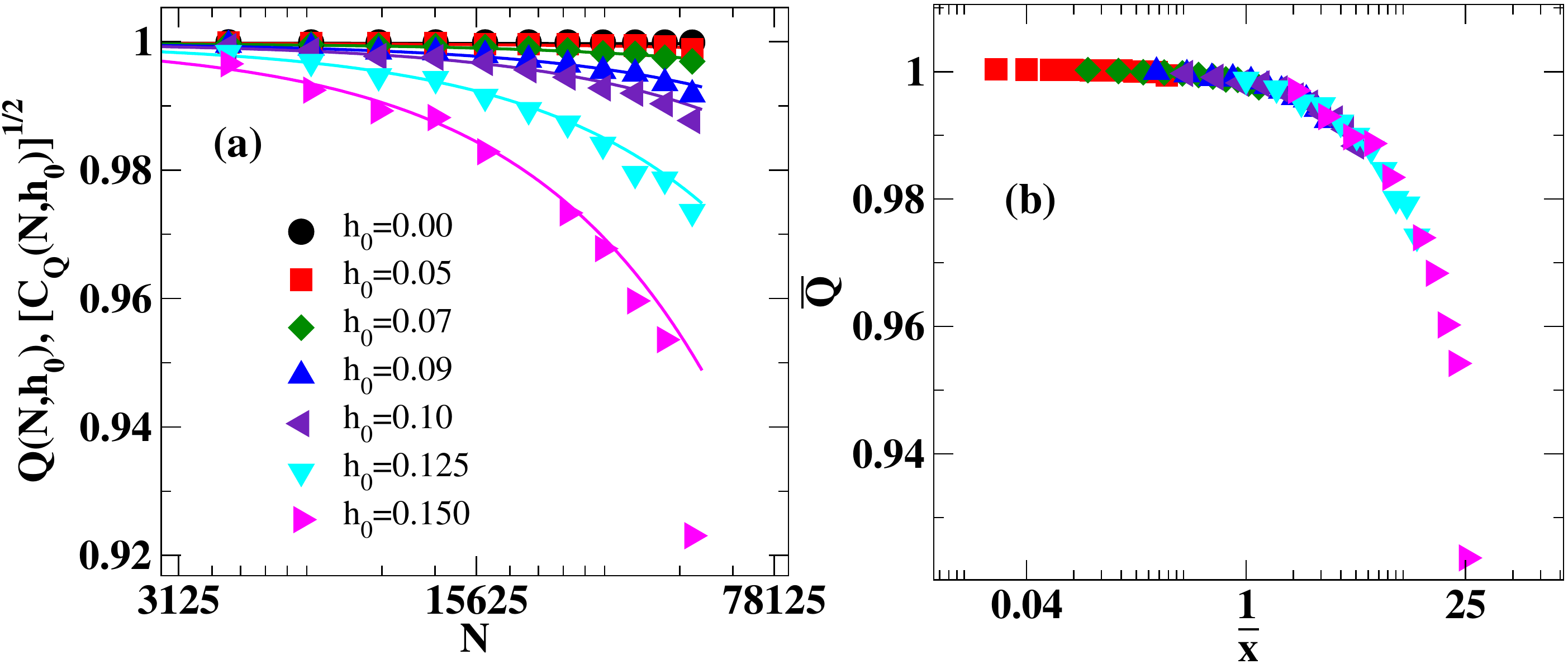}}   
\mbox{\includegraphics[width=4 cm,height=3cm]{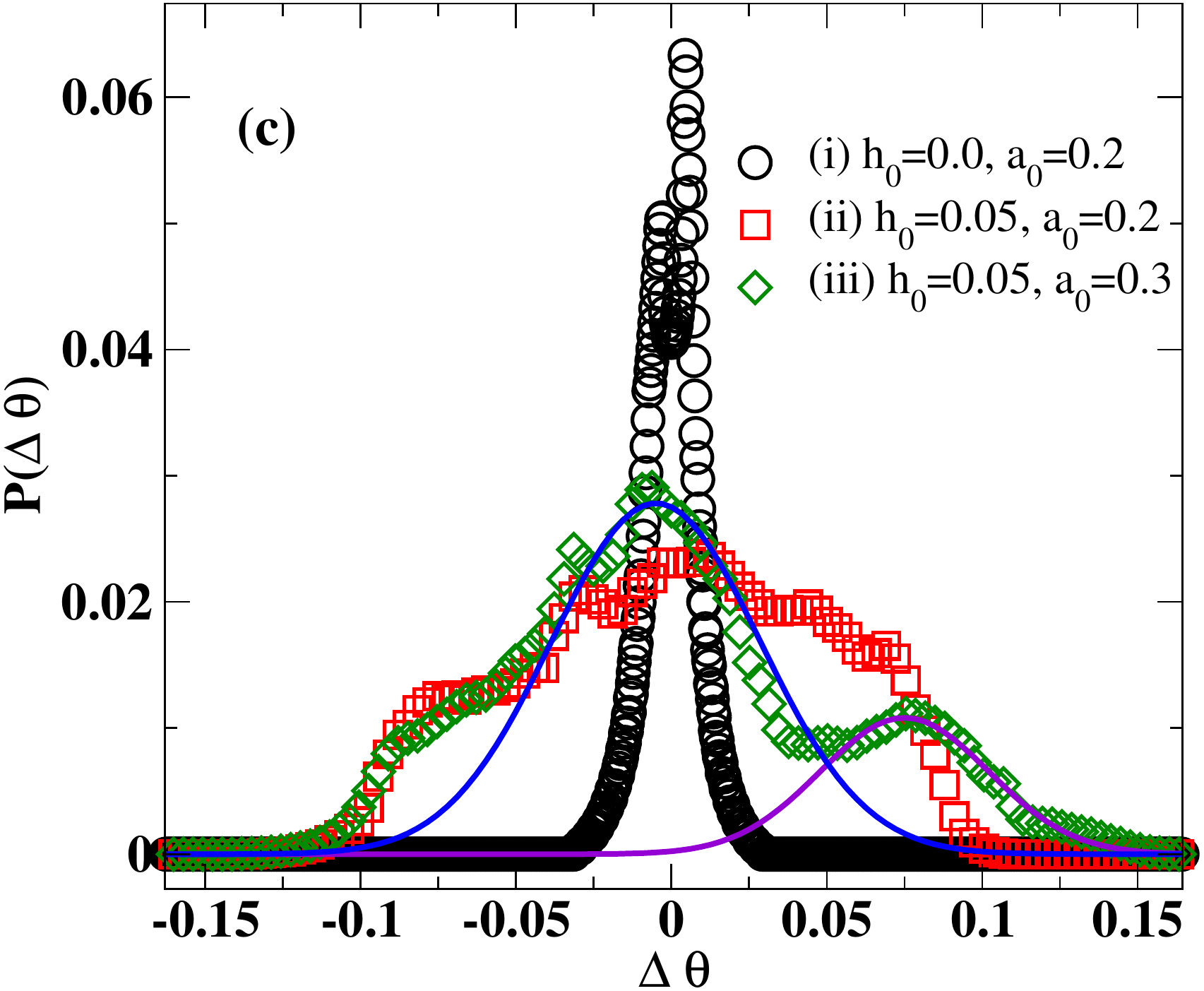}}

\caption{(Color online) (a) NOP  ${\bf Q}$ {\em vs.} system size $N = K \times K$ for different $h_0$ (symbols) 
and $\sqrt{C_{\mathcal{Q}}}$, Eq. (\ref{eq: 3}) (solid lines), 
(b) Scaled NOP, $\bar{Q}=Q \times N^{\mathcal{B'}/2}$ {\em vs.}  $\bar{x}=   N \times {h_0}^{4}$.
 (c) Probability distribution function $P(\Delta \theta)$ of angle fluctuations of angular orientation for different values of $(h_0,a_0)$ i.e. $(0.0, 0.2)$ ,$ (0.05, 0.2)$ and $(0.05, 0.3)$. Data for  $h_0=0.05, \ a_0=0.3$ (data points) fitted with Gaussian for two distinct peaks (solid lines).}
\label{fig: 2}
\end{figure*}

We also studied the effect of disorder on the dynamics of the defects and the ordering kinetics.
The {\em effective} dynamic growth exponent, $z_{eff}$ \cite{bray1994}
increases on increasing disorder. The two-point correlation functions for both fields show good dynamics scaling for all disorder, 
but no static scaling is found for different disorder.


We construct a monolayer of self-propelled apolar particles of length $l$, 
on a two-dimensional substrate of friction coefficient $\chi$. 
Each particle is driven by an inherent driving force $ F $ acting along the particle's long axis. 
The ratio of the force $F$ to the friction coefficient gives a constant self-propulsion speed $v_0 = F/\chi$ to each particle. 
The apolar nature of the particle makes them move forward and backward with {\em equal} probability 
with a step size equal to $v_0$.
On a time scale, large compared to the interaction time and length scale much larger than the particle size, the dynamics of the system is governed by coupled hydrodynamic equations of motion for slow variables {\em viz.} local density $\rho({\bf r}, t)$, and local NOP $\mathcal{Q}({\bf {r}}, t)$ \cite{pggennes1993}, 


\begin{equation}
\centerline{$\partial_{t} \rho=a_{0} \nabla_{i}\nabla_{j}\rho \mathcal{Q}_{ij} + D{\rho}   \nabla^2 \rho$}
\label{eq: 1}
\end{equation}

\begin{widetext}
\begin{equation}
	\centerline{$\partial_{t}\mathcal{Q}_{ij}=[\alpha_{1}(\rho)-\alpha_{2}(\mathcal{Q} : \mathcal{Q})]\mathcal{Q}_{ij} +\beta(\nabla_{i}\nabla_{j} - \frac{1}{2}\delta_{ij}\nabla^2) \rho + D_{\mathcal{Q}}\nabla^2 \mathcal{Q}_{ij}+H_{ij} + \Omega_{ij}$} 
\label{eq: 2}
\end{equation}
\end{widetext}

The  Eqs. (\ref{eq: 1}) and (\ref{eq: 2}) written in dimensionless units by rescaling all lengths by the length  of the 
particle and time  by the collision time and are of the same form as derived
from the microscopic rule-based model in \cite{shradhanjop}, with an additional term due to 
{\em quenched disorder}. 
The quenched disorder is introduced  as {\em random field } in the free energy density $\mathcal{F} = -{\bf \mathcal{Q}} : ({\bf h}{\bf h}-\frac{{\bf I}}{2})$.
Which further leads to  $ H_{ij}  = (h_ih_j-h_0^2\frac{1}{2}\delta_{ij})$,
in equation \ref{eq: 2}, in two-dimensions $i,j=1,2$ are the spatial indices for the two components of vectors. Where,  $h_{i}=h_0(cos\phi,sin\phi)$, here $h_{0}$ is the disorder strength and $\phi({\bf r})$ is
a uniform random angle between $(0, 2\pi)$, 
with mean zero, quenched in time (no time dependence) and space correlation  $\langle \phi({\bf{r}}) \phi({\bf{r'}})\rangle= \delta ({\bf{r}}-{\bf{r'}})$.

The last term, $\Omega_{ij}$ is a tensorial symmetric traceless white noise with mean zero, such that $\langle \Omega_{ij}({\bf{r}},t) \Omega_{kl}({\bf{r'}},t')\rangle= \bigtriangleup_0  \delta ({\bf{r}}-{\bf{r'}}) \delta(t-t')\epsilon_{ijkl}$. Here, $\bigtriangleup_0$ is the noise strength and $\epsilon_{ijkl}=\frac{1}{2}(\delta_{ik}\delta_{jl}+\delta_{il}\delta_{jk}-\delta_{ij}\delta_{kl})$. \\

In the above Eqs. \ref{eq: 1} and \ref{eq: 2} we keep the model minimal and ignore
the flow field \cite{lgiomiprl2013}, completely; or assume the interaction among the particle is short range
volume exclusion, and no hydrodynamic interaction.  Hence the system we study is {\it dry} active nematic. 

\begin{figure}
\includegraphics[width=4.5 cm,height=3.5 cm]{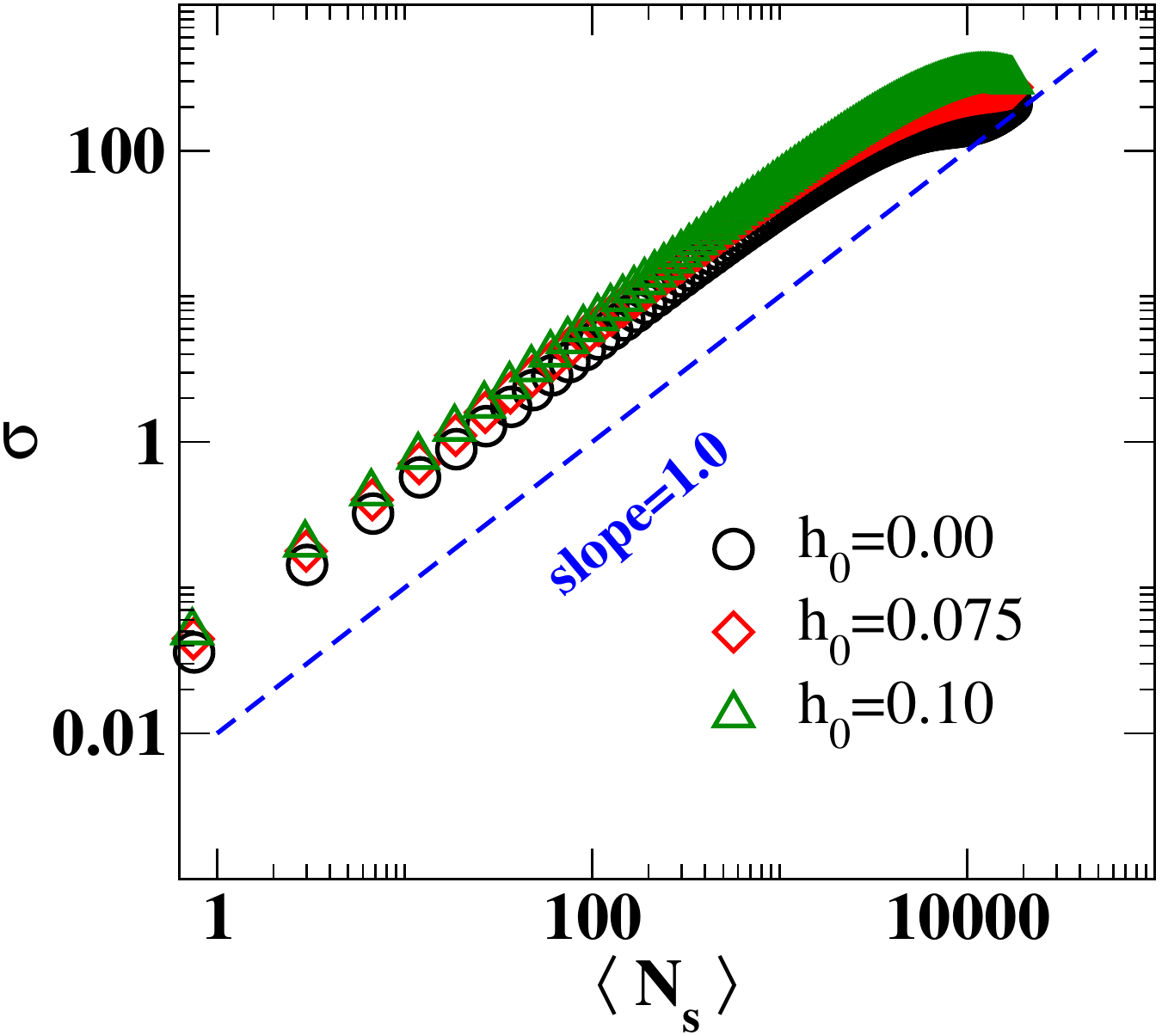}
\caption{(Color online) Number fluctuation $\sigma$ {\em vs.} $<{N_s}>$ plot for different disorder strengths, and $K=300$.}
\label{fig: 3}
\end{figure}

The random field introduced in our current model is similar to the random field
 in XY-model (RFXY-model) \cite{imryma1975}. 
Hereafter we refer our model as random field active nematic (RFAN) when $h_0\ne 0$, and  clean-active nematic (clean-AN) for $h_0=0$.\\
To perform the numerical integration of Eqs. \ref{eq: 1} and \ref{eq: 2} we construct a  two-dimensional $K \times K$ square lattice 
with periodic boundary condition (PBC) and discretise the space and time derivatives   
using Euler scheme ($\Delta x = 1.0$ and $\Delta t = 0.1$). Initially, we start with random 
homogeneous  density, with mean ($\rho_0=0.75$), and random  orientation. \\
We first study the steady-state of the system for $h_0 = (0.0, 0.15)$ and system size, i.e. $K= 64 \ to \ 512$. Coarsening is studied for larger $K=1024$. 
Steady-state results are obtained for simulation time $t=\mathcal{O}(10^6)$ and 
the average over $10$ independent realisations. 
 One simulation time is counted after update of Eqs. \ref{eq: 1} and \ref{eq: 2} for all
lattice points.
Parameters in Eqs. (\ref{eq: 1}) and (\ref{eq: 2}) are ($a_0=0.1 \ to \ 0.3$), $D_{\rho}=\frac{1}{4}(a_0^2+1)$, $ \rho_0=0.75, \ \rho_c=0.5, \ \alpha_2=1,  
\ \beta=0.25,  \ and \ D_{\mathcal{Q}}=1$, $\bigtriangleup_{0} =10^{-4}$ and we check that, system remains stable for the chosen set of parameters.\\

{\em Results}:- 
We first measure the steady-state properties of RFAN for different $h_0$.
The global ordering in the system is measured by  calculating the 
nematic order parameter (NOP) defined as 
${\bf Q}=\langle \frac{1}{N}\sqrt{\vert \sum_{i=1}^{N}cos (2\theta_i)\vert^2+\vert \sum_{i=1}^{N}sin (2\theta_i)\vert^2}\rangle $, 
where the sum runs over all the lattice points.
$\langle .. \rangle$ shows the average over many realisations. 
We compare the measured ${\bf Q}$ in numerical simulation with the analytical expression for the two-point orientation correlation function $C_{\mathcal{Q}}$ (Eq. \ref{eq: 3}),
 obtained from a linearised treatment of small 
fluctuation in a uniform ordered phase. 
We calculate the equal time Fourier transformed spatial correlation of 
angle, $S_{{\bf q}}(\theta)$ as a function of wavevector ${\bf q}$ and present our result for 
$q_x = q_y$. Starting from the equations of motion \ref{eq: 1} and \ref{eq: 2}, a straight forward 
linearised approximation shows that for a finite disorder strength,

\begin{equation}
\centerline{$C_{\mathcal{Q}}(N) \simeq \frac{1}{N^{\mathcal{B'}}} e^{-\mathcal{C'}h_{0}^{4} N}$}
\label{eq: 3}
\end{equation}

which is obtained from the inverse Fourier transform of orientation
structure factor $S_{{\bf q}}(\theta)$ at wave-number $q \simeq N^{-1/2}$. 
The coefficients in Eq (\ref{eq: 3}), $\mathcal{B}'=1.17 \times 10^{-4}$ and $\mathcal{C'}=3.9 \times 10^{-3}$ (for $a_0=0.2$) are constants and depends on the system parameters.  
Hence ${C_{\bf Q}}$ is a product of algebraic and exponential decay with $N$. 
Fig. \ref{fig: 2}(a), data points shows the   plot of NOP {\em vs.}
 $N$ for different disorder strengths, $h_0$. 
For clean-AN, ${\bf Q}$ decays algebraically as $N^{-B}$, where  $B \simeq 1.05 \times 10^{-4}$, depends on system parameters. Also, $B$ is small, since, for the given parameters, clean-AN is in the deep ordered state and significantly away from the isotropic-nematic transition, 
where it shows the instability \cite{shima2011,  rakeshscrep, Shi2013, Shi2014}.
 For finite disorder, ${\bf Q}$ shows the deviation from the pure algebraic decay. It decays algebraically for small $N$
and leads a crossover to exponential decay for a larger $N$. The larger the disorder strength, the crossover to exponential decay appears for smaller $N$.

\begin{figure*}      
    \mbox{\includegraphics[width=7cm,height=3. cm]{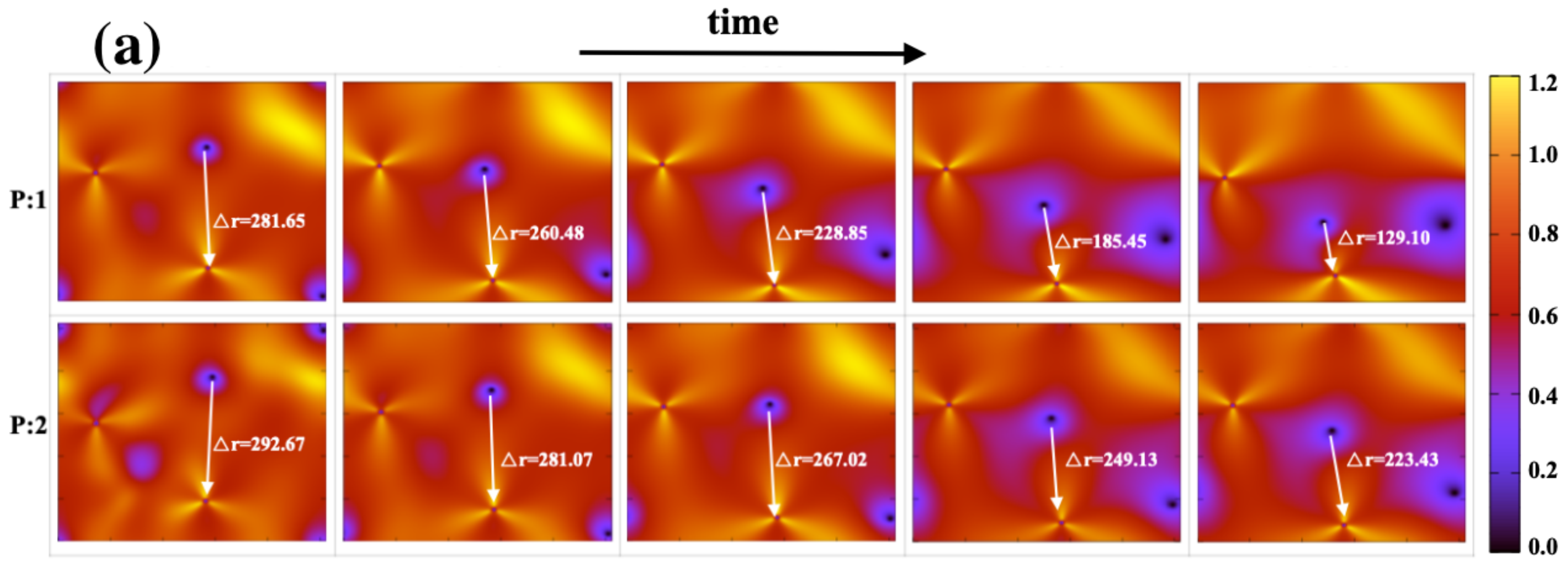}}   
    \hspace{1.0px}
    \mbox{\includegraphics[width=3.6 cm,height=2.6 cm]{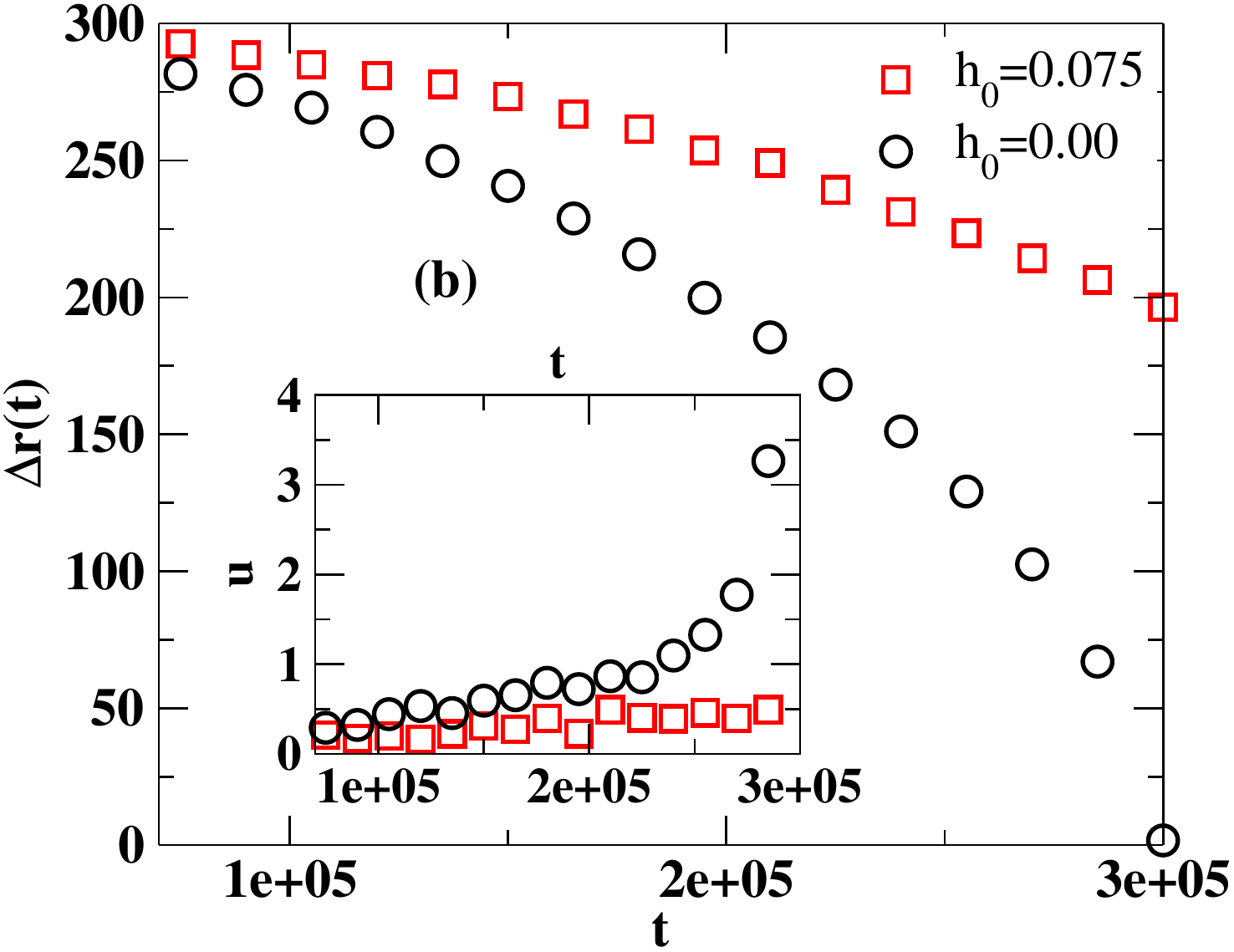}}
    \hspace{1.0px}
    \mbox{\includegraphics[width=6 cm,height=2.5 cm]{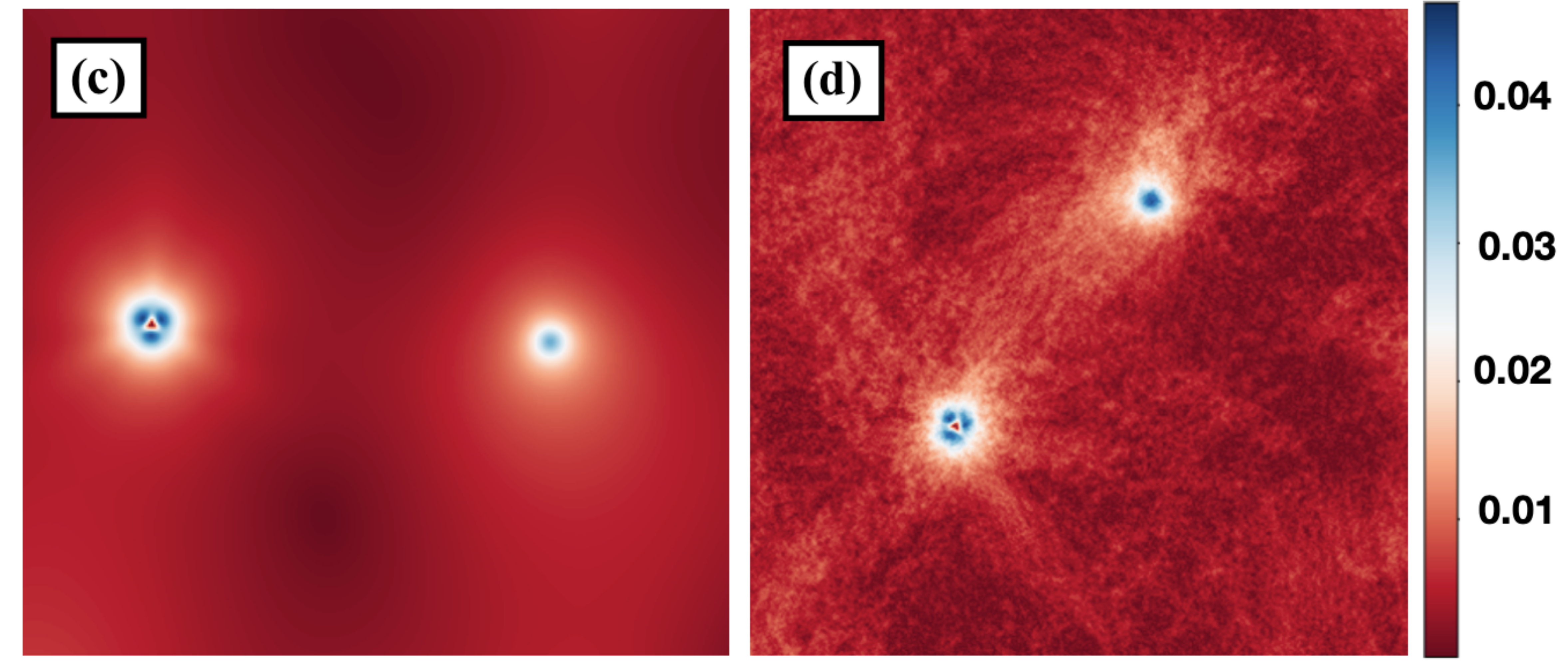}}

\caption{(Color online) (a) Snapshots of local NOP $\mathcal{Q}$: upper panel (P:1) is for clean-AN (i.e. $h_0=0.0$) and bottom panel (P:2) is for RFAN ( $h_0=0.075$) and the number along the white arrow is the  relative separation $(\Delta r(t))$ between the $+$ and $-1/2$ defects. (b) $\Delta r(t)$ {\em vs.} $t$ plot for  $h_0=0.0, 0.075$.  $u$ is the relative speed of defects defined as $u=|\frac{d}{dt}\Delta r(t)|\times 10^{-3}$ and plotted in the inset of (b).  (c) Snapshot of density current  near the defects for clean-AN, $h_0=0.0$ and (d) RFAN, $h_0=0.075$. Intensity of colors shows the magnitude of the density current. Data is generated for system size $N = 512^2$.}
\label{fig: 4}
\end{figure*}

In Fig. \ref{fig: 2}(a) lines 
are plot of $\sqrt{C_\mathcal{Q}(N, h_0)}$, Eq. (\ref{eq: 3}). 
For clean-AN, $C_{\mathcal{Q}}(N, 0)$ is pure power-law whereas, for RFAN, it decays exponentially for higher $N$. 
Hence crossover happens after a disorder dependent $N$, $N_c(h_0) \sim h_0^{-4}$.  
We find a good match of lines (eq. \ref{eq: 3})  and data from the simulation.  We see a systematic deviation between data points from the simulation and the linear study lines for large $h_0$, which is due to nonlinearities present in the model. 
In Fig. \ref{fig: 2}(b) we plot the ${\bf Q}\times N^{\mathcal{B'}/2}$ {\em vs.} 
$N \times h_0^4$ for different $h_0$ and find a good collapse of data for different disorder strengths.\\
To further understand the steady state in RFAN, we calculate the probability distribution function (PDF) $P(\Delta \theta)$ of angle fluctuations  $\Delta \theta$ 
from the mean direction. Fig. \ref{fig: 2}(c) shows the  plot of $P(\Delta \theta)$ {\em vs.}
$\Delta \theta$ for different $(h_0, a_0)$. 
$P(\Delta \theta)$ for clean-AN shows a very narrow peak at $\Delta \theta=0$, 
whereas, for RFAN,  PDF has a much broader distribution and more than one peak at non-zero $\Delta \theta$ (see the appendix \ref{appendix}  for snapshots).  Lines in Fig. \ref{fig: 2}(c) are fit to two distinct peaks 
for $h_0=0.05, a_0=0.3$  with Gaussian. On increasing activity, the width of the distribution sharpens, and more distinct peaks emerge. Hence it infers the stronger intra-domain ordering and distinct ordered domains for large activity.
Similarly, for a more considerable disorder, $P(\Delta \theta)$ shows a more number of such different peaks, which means that smaller domains emerge more if we further increase the disorder strength (data not shown).

In the  appendix \ref{appendix} and SM ( see the supplementary materiel for animations), we show the animation for the snapshots of local NOP for clean, $h_0=0.05 $ and $h_0=0.1$, which shows that for a clean system, the final state is globally ordered, whereas, for RFAN, different ordered domains are formed and survived at late times.

We also calculate the steady-state density fluctuation, for all disorder, 
number fluctuation $\sigma = \sqrt{\langle {N_s}^2 \rangle -\langle {N_s} \rangle^2} \sim <{{N_s}}>$, where ${N_s}$ is the mean number of particle in subcells, fig. \ref{fig: 3}, 
which shows a giant number fluctuation (GNF), as found in \cite{shradhaprl, husechateprl2004, sriramadititonerepl2000}. 
 When compared with the linearised calculation of two-point density structure factor, as given in Eq. (\ref{eq: 42}), for $q \simeq N^{-1/2}$, 
density fluctuation should show the fluctuations larger than the clean 
AN. But large fluctuation can arise for size ${N_s} >$ ${N}_c$ $\sim h_0^{-4}$, which is hard 
to achieve in numerical simulation. Hence in general, inhomogeneity does not affect the density
fluctuations in active nematic, although it significantly changes the nature of two-point 
orientation correlation function. 
Now we further study the ordering kinetics to such steady-state. \\

\begin{figure}      
\mbox{\includegraphics[width=7.cm,height=2.6cm]{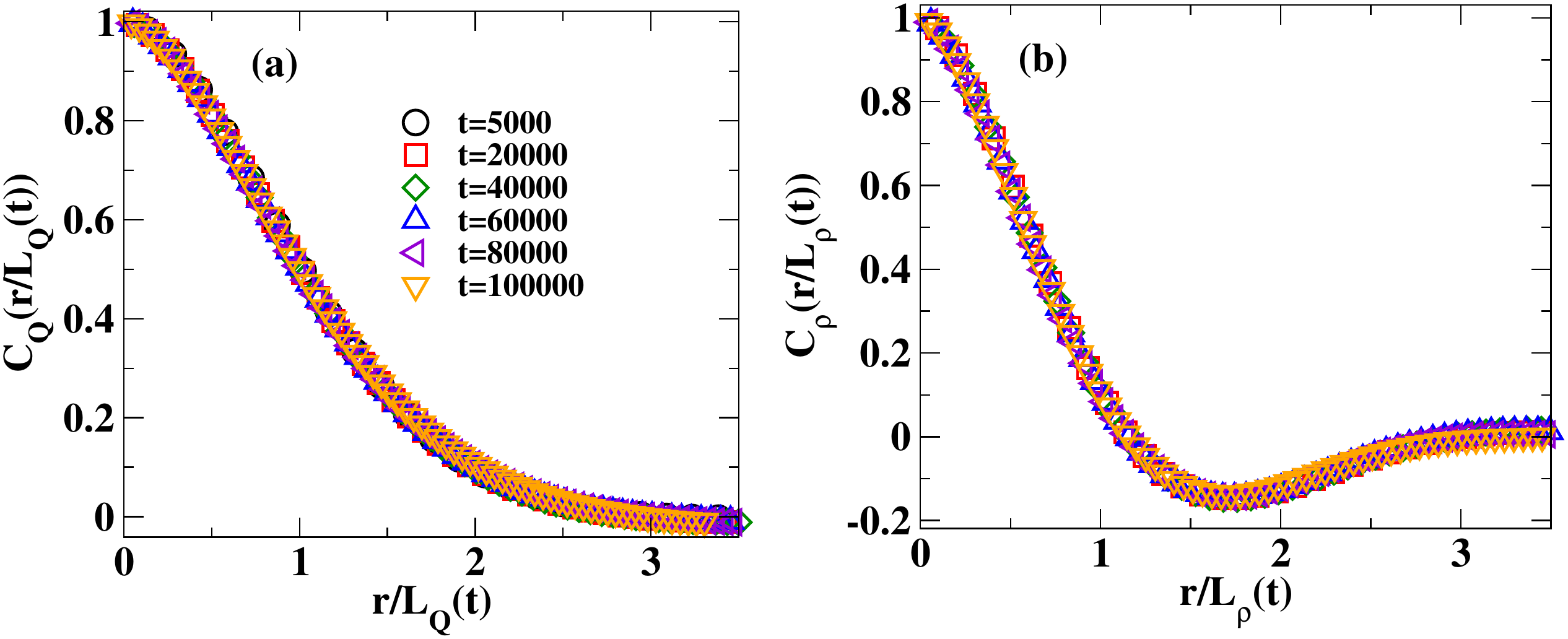}}   
\hspace{1.0px}
\mbox{\includegraphics[width=7.cm,height=2.6cm]{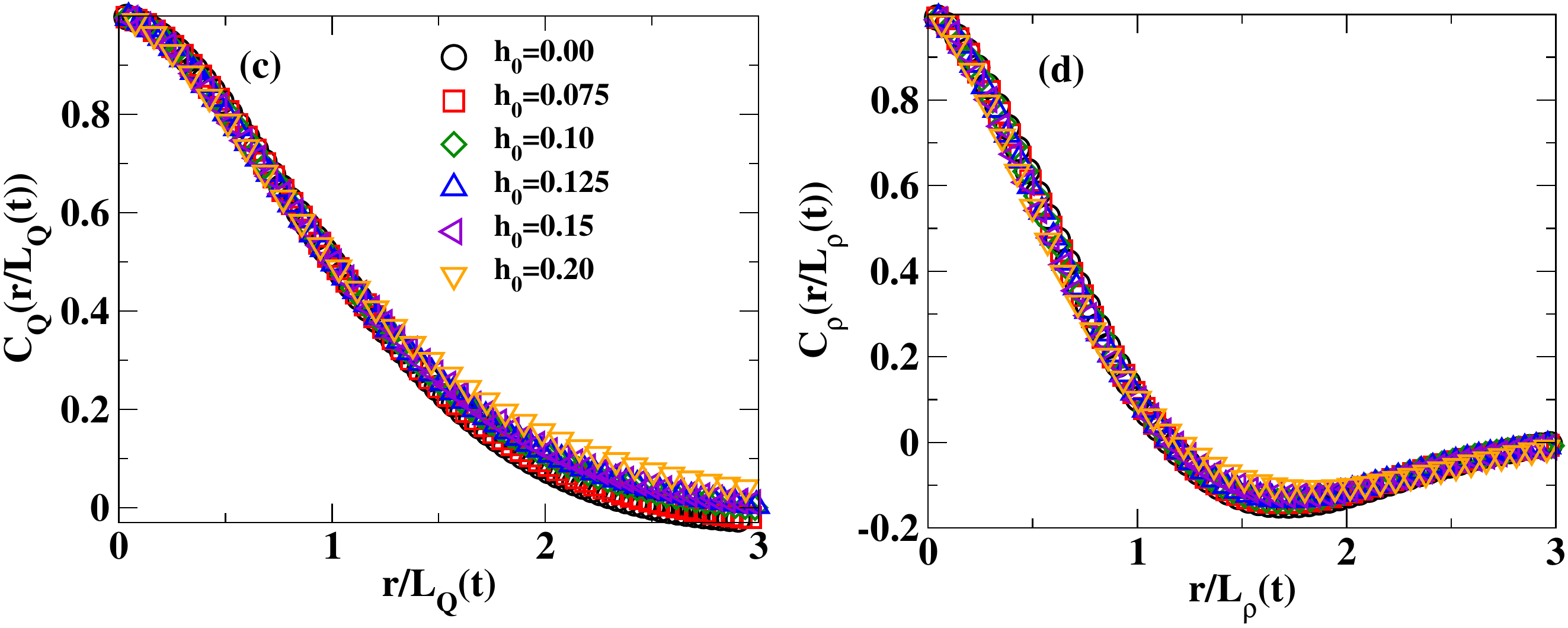}}
\caption{(Color online) Two-point correlation function  $C_{\mathcal{Q}, \rho}$ {\em vs.} scaled distance $r/L_{{\mathcal{Q}, \rho}}(t)$.  (a-b)  Two-point correlation function for RFAN i.e. $h_0=0.1$ at different simulation time ($t$).  (c-d)  Two-point correlation function for different $h_0$ and at fixed simulation time $t=10^5$.}
\label{fig: 5}
\end{figure}

\begin{figure}
\centering
\mbox{\includegraphics[width=6cm,height=2.5 cm]{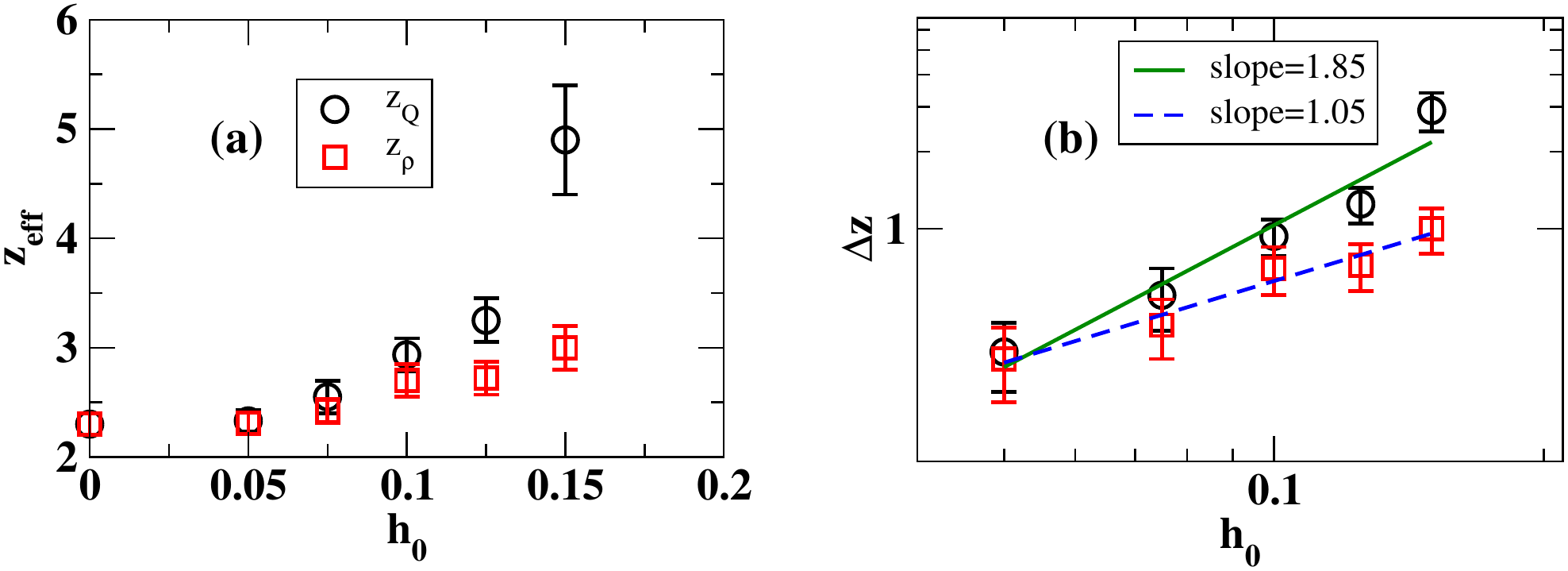}	}
\caption{(Color online) Plots of dynamic growth exponent $z_{eff}(h_0)$  {\em vs.} disorder strength $h_0$ (a), $\Delta z$ vs. $h_0$ on $\log - \log$ scale (b).}
\label{fig: 6}
\end{figure}

{\em Kinetics:-}  

When the system  brought from a disordered state to an
ordered state,  ordering happens through the process of domain formation and which is due to
the creation and annihilation of   $\pm 1/2$  topological defects.

In two-dimensional active nematic, these defects have topological geometry \cite{defecttopology}. 
A  $ +1/2 $ defect has a comet-like structure and moves along the axis parallel to its tail, whereas a $-1/2 $ defect has a three-fold symmetry and does not have any preferred direction of motion \cite{vnarayan, shradhaphystrans, lgiomiprl2013}.  
The dynamics of defects play a vital role in the ordering of the system \cite{bray1994}. 
In Fig. \ref{fig: 4}(a-b), we study the dynamics of defects for a clean-AN ($h_0=0.0$) as well for RFAN  ($h_0=0.075$). 
Fig. \ref{fig: 4}(a)  shows the snapshots of local NOP, 
$\mathcal{Q}$, for clean-AN (upper panel) and RFAN  (lower panel). 
White arrows show the relative separation, 
$\Delta r(t)$, between a pair of $\pm 1/2$ defects. The arrow's tail and head represent the position of $+1/2$ and $-1/2$ defects, respectively. We see that the disorder slows the dynamics of the  $+1/2$.
The variation of  $\Delta r(t)$ {\em vs.} $time$ is shown in Fig. \ref{fig: 4}(b). 
The length of the white arrow in Fig. \ref{fig: 4}(a) decreases with time 
(or $\Delta r$ decay with time Fig. \ref{fig: 4}(b)), which shows the two
defects come close to each other.  
For clean-AN, $\Delta r(t)$ decay at a faster rate, whereas it takes a longer time in the presence of disorder.  
Hence the relative speed is small in the presence of disorder, as shown in the inset of Fig. \ref{fig: 4}(b). 
To further understand the mechanism of slowing down of defect dynamics,  in Fig. \ref{fig: 4}(c-d),  we show 
 the snapshot of the local density current   near a pair of  $\pm 1/2$ defects. Density current ${\bf J_{\rho}}$ defined from Eq. (\ref{eq: 1}), which can be rewritten as continuity equation, $\partial_t \rho = -\nabla \cdot {\bf J_{\rho}}$ where ${\bf J_{\rho}}= -a_0 \nabla \cdot ( \rho \mathcal{Q})-D_{\rho} \nabla \rho$. 
The intensity of colors shows the magnitude of the density current. 
For clean-AN, current flow is smooth near the defects  Fig. \ref{fig: 4}(c), whereas with  disorder ($h_0=0.075$), 
current flow is distorted, Fig. \ref{fig: 4}(d), which results in slower growth dynamics, we will discuss next.
We also studied the effect of activity on the relative speed of a pair of defects. For larger activity, the relative separation (starting from
the same relative separation)
between a pair of defects decreases faster, and hence they annihilate quickly. Which, in turn, results in more ordering for the same disorder strength (see the appendix \ref{appendix} for details). \\

{\em Growth law and scaling properties}
As we discussed in previous paragraph, disorder affect the defect dynamics, and it can further influence the kinetics of domain
ordering. We characterise the domain growth by calculating the correlation functions 
for orientation  $\mathcal{Q}$ , $C_{\mathcal{Q}}({\bf{r}},t)=\langle\mathcal{ Q}({\bf{0}},t):\mathcal{Q}({\bf{r}},t)\rangle$ and, local density $\rho$, $C_{\rho}({\bf{r}},t)=\langle \delta \rho({\bf{0}},t) \delta \rho ({\bf{r}},t)\rangle$, 
where $\delta \rho({\bf{r}},t)=\rho({\bf{r}},t)-\rho_0$ is the deviation of the local density 
from the mean $\rho_0$.
With time both 
correlations increases due to domain growth. 
Fig. \ref{fig: 5}(a-b) show the plot of $C_{\mathcal{Q}}(r/L(t))$ and $C_{\rho}(r/L(t))$ {\em vs.} 
scaled  distance $r/L_{\mathcal{Q}, \rho}(t)$ and they all collapse to a single curve. Where the characteristic length 
$L_{\mathcal{Q}, \rho}(t)$ is calculated from the first zero crossing of  
 $C_{\mathcal{Q}}({\bf{r}},t))$ and $C_{\rho}({\bf{r}},t)$. 
Fig. \ref{fig: 5}(c-d) shows the plot of $C_{\mathcal{Q}}(r/L_{\mathcal{Q}}(t))$ and $C_{\rho}(r/L_{\rho}(t))$ {\em vs.} 
scaled  distance $r/L_{\mathcal{Q}, \rho}(t)$ calculated at equal  time ($t=10^5$) for different disorder $h_0$.
We find no scaling for different disorder strengths for both $\mathcal{Q}$ and $\rho$.
Therefore, for all disorder strengths, the system shows good dynamic scaling but no static scaling in orientation and density. \\
The equilibrium analogue of clean-AN is XY-model and the characteristic length of growing domain in two-dimensional 
$XY-$model goes as $L_{XY}(t) \sim (t/\ln(t))^{1/2}$ \cite{bray1994, pargellis1992}. 
Hence we assume that for RFAN, $L_{\mathcal{Q}, \rho} \sim (t/\ln(t))^{1/z_{eff, \mathcal{Q}, \rho}}$ and  further  calculate the dynamics growth 
exponent $z_{eff,\mathcal{Q},\rho}$ from correlation length $L_{\mathcal{Q}, \rho}(t)$, 
defined as $\frac{1}{z_{eff,  \mathcal{Q}, \rho}} =\langle \frac{d \ln L_{\mathcal{Q}, \rho}(t)}{d (t/\ln t)} \rangle$ \cite{pargellis1992, yurke1993, bray1994, shradhaphystrans}, where $\langle .. \rangle$ is mean 
value of $z_{eff}$ over intermediate time ({\bf  $ t \sim 1000 \ to \ 15000 $}) when it remains constant for at least one decade, fig. \ref{fig: 6}(a). 
We find that $z_{eff,\mathcal{Q},\rho} \simeq 2$ for Clean-AN
and increases on increasing $h_0$. In fig. \ref{fig: 6}(b), we  plot the $\Delta z= z_{eff}-2$ vs. $h_0$ on $\log-\log$ scale.
The change $\Delta z$ increases algebraically with $h_0$ with power $\sim$ $2$ and $\sim 1$ for $\mathcal{Q}$ and $\rho$ respectively. Hence growth kinetics of density field shows 
small change in comparison to orientation field. Or small change in growth kinetics of density field affects the orientation field substantially.\\

\begin{figure}
\includegraphics[height=3cm, width=7cm]{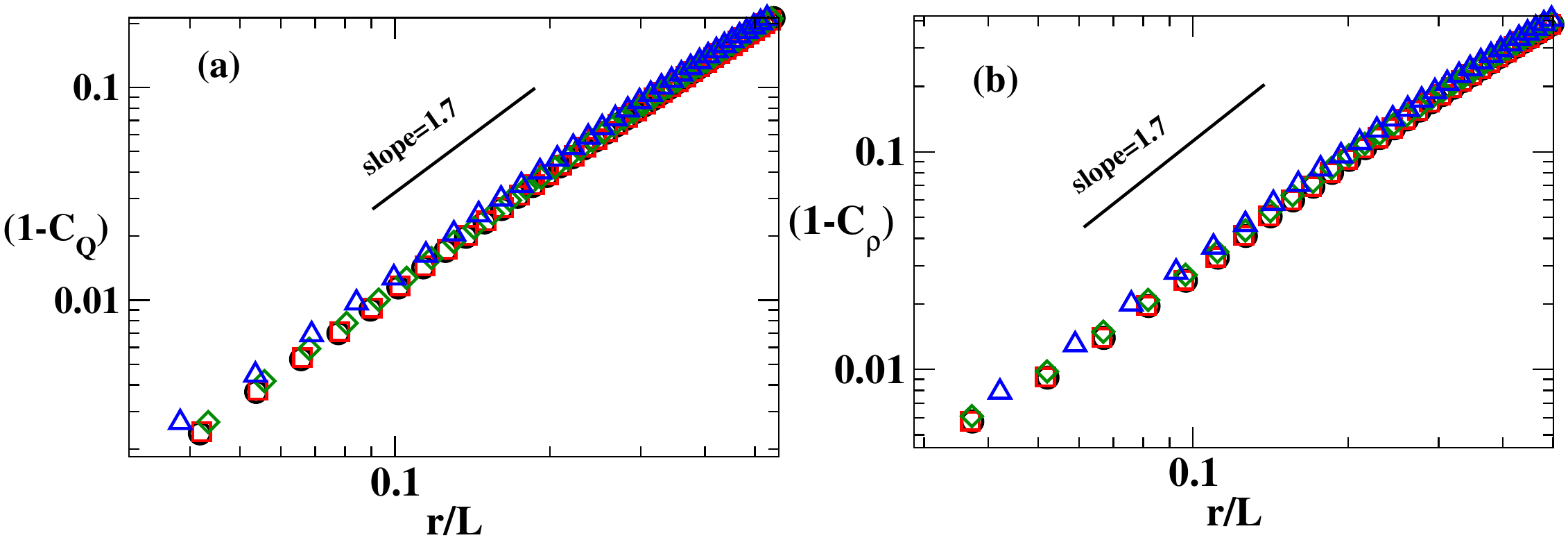}
\caption{(Color online) Cusp's exponent $\alpha$ (slope of the plot) for $Q$ and $\rho$ field. Diffrent symbols used for diffrent values of dissorder strength: $h_0=0.0$ (circle), $h_0=0.05$ (square), $h_0=0.075$ (diamond), $h_0=0.10$ (triangle).}
\label{fig: 7}
\end{figure}

{\em Morphology of ordered domains} 
We  study the effect of disorder on the morphology of ordering domains.  We calculate  
 the behaviour of scaled two-point correlation functions $C_{\mathcal{Q}, \rho}(r/L_{\mathcal{Q}, \rho})$  for small
$r/L_{\mathcal{Q}, \rho}$. In the limit of small $r/L_{\mathcal{Q}, \rho}$, 
$C_{\mathcal{Q}, \rho}(r/L_{\mathcal{Q}, \rho})   \sim 1-(\frac{r}{L_{\mathcal{Q}, \rho}})^{\alpha}$, where $\alpha$ is called the 
cusp exponent and features the domain morphology \cite{bray1994, Mbarma2000}. 
In Fig. \ref{fig: 7} we plot the $1-C_{\mathcal{Q}, \rho}(r/L_{\mathcal{Q}, \rho})$ {\em vs.} scaled distance $r/L_{\mathcal{Q}, \rho}$ on 
$\log-\log$ scale and estimate the cusp exponent $\alpha$ for both fields  $(\mathcal{Q}, \rho)$. 
The exponent, $\alpha \simeq 1.7$ for both fields and for all
disorder strengths. Hence domain morphology remains unaffected in the presence of disorder.\\

{\em Discussion:-} 
We studied two-dimensional dry active nematics with the quenched random disorder using the
  the hydrodynamic equations of motion for the slow fields, {\em viz.} density 
$\rho$ and orientation $\mathcal{Q}$, in a coarse-grained description.
 
The study from the numerical solution of equations of motion and the linearized hydrodynamic calculation 
shows that the orientation correlation follows a crossover from QLRO (algebraic decay of correlation ) to 
SRO (exponential decay). Such crossover occurs due to the pinning of $\pm 1/2$ defects in the presence of finite disorder, which breaks the system in domains of different orientations. The size of such domains decreases on increasing disorder.  For clean as well as RFAN, number fluctuation is giant. \\
We also studied the approach to the steady-state by (i) characterizing the dynamics of $\pm 1/2$ defects and (ii) calculation of the characteristic length of growing domains  $L_{\mathcal{Q}, \rho}(t)$. 
The slow dynamics of $+1/2$ defect leads to the slower domain growth in the presence of disorder. 
Although domain growth is slower in the presence of disorder, the two-point correlation function for both fields $C_{\rho, \mathcal{Q}}$ shows good dynamic scaling. Still, no static scaling is found for
different disorder strengths. Domain morphology remains unaffected in the presence of disorder.\\

We find an interesting steady-state in RFAN, which is different from its corresponding equilibrium counterpart: random 
field XY-model \cite{imryma1975}.
 Our study should motivate experimentalists to verify our findings and encourage us to study the effect of other kinds of disorders in active nematics. 
To make the model minimal, the effect of background fluid is ignored in our present study; hence it is for dry active nematic. It would be interesting to extend
this study for wet active systems \cite{lgiomiprl2011, lgiomiiop2012, lgiomiprl2013}.

{ \em Acknowledgements: }
SM would like to thank Sriram Ramaswamy and Sanjay Puri for useful discussion at the beginning of the project. SK would like to thank Manoranjan Kumar, Debprasad Giri and Rajeev Singh for useful suggestions. SK thanks IIT(BHU) Varanasi and SNBNCBS Kolkata for computational facility. SM and SK, thanks DST-SERB India, ECR/2017/000659 for  financial support. 

\bibliographystyle{apsrev4-1}
\bibliography{citations}

\appendix
\section{Linearised hydrodynamic calculation of two-point correlation functions}
\label{analytics}

We start with the hydrodynamic equations of motion for local density $\rho$ and NOP $\mathcal{Q}$ as introduced in Eq. (\ref{eq: 1}) and (\ref{eq: 2}),

First, two terms on the R. H. S. of Eq. (\ref{eq: 2}) are the mean-field terms:  
where, $\alpha_{1}(\rho)=\alpha_0 (\frac{\rho}{\rho_{c}}-1)$ where $\rho_{c}$ is the critical 
density, where $\alpha_0=1$   is chosen as unity for simplicity.
System shows homogeneous ordered state for $\alpha_{1}(\rho_{0})>0$, and disordered isotropic state when $\alpha_{1}(\rho_{0})<0$,
where $\rho_0$ is the mean density of particles.
The third term is coupling to the density field 
and the fourth term is the diffusion in $\mathcal{Q}$. Origin of such
diffusion  can be obtained from 
the equal elastic constant approximation of Frank-free energy for two-dimensional
equilibrium nematic \cite{chaikinlubenskey, pgdegennes}.

We rewrite Eqs. (\ref{eq: 1}) and (\ref{eq: 2}) neglecting the higher order fluctuations about the homogeneous ordered  steady state. 
The local nematic order parameter $\mathcal{Q}$ is given as $\mathcal{Q} =\frac{S}{2} \; \left[ \begin{array}{cc} \cos 2\theta & \sin 2\theta \\ \sin 2\theta & -\cos 2\theta  \end{array} \right]$ where, $S$ is a scalar and a measure of ordering. We define $\delta \rho$, $\delta S$ and $\theta$ as the fluctuation terms from their mean values $\rho_0$, $S_0$ and $\theta_0$ respectively. Here, $S_0=\sqrt{\frac{2\alpha_1(\rho_0)}{\alpha_2}}$, and is obtained from (Eq. (2) main text) for homogeneous steady state. Therefore, to linear order we have $\mathcal{Q}_{11}=\frac{1}{2}(S_0+\delta S)$,  $\mathcal{Q}_{12}=\theta S_0$, and (Eq. (1) main text) gives,

\begin{widetext}

\begin{equation}
\centerline{$\partial_t(\rho_0+\delta \rho)=a_0 [{\partial_x}^2  {(\rho_0+\delta \rho)\frac{(S_0 +\delta S)}{2}}+{\partial_y}^2 {(\rho_0+\delta \rho)\frac{(-(S_0 +\delta S))}{2}}+2\partial_x \partial_y {(\rho_0+\delta \rho)S_0 \theta}]+D_{\rho} \nabla^2 (\rho_0+\delta \rho)$}
\label{eq: 18}
\end{equation}

or,

\begin{equation}
\centerline{$\partial_t \delta \rho=a_0 [(\partial_{x}^2 \frac{(S_0 \delta \rho+\rho_0 \delta S)}{2}-\partial_{y}^2\frac{(S_0 \delta \rho+\rho_0 \delta S)}{2}+2S_0 \rho_0 \partial_x \partial_y \theta]+D_{\rho} (\partial_{x}^2+\partial_{y}^2)\delta \rho$}
\label{eq: 19}
\end{equation}
or,
\begin{equation}
\centerline{$\partial_t \delta \rho=(\frac{a_0S_0}{2}+D_{\rho})\partial_{x}^2 \delta \rho +(D_{\rho}-\frac{a_0S_0}{2})\partial_{y}^2 \delta \rho +\frac{a_0\rho_0}{2}(\partial_{x}^2-\partial_{y}^2)\delta S+2a_0\rho_0 S_0\partial_x \partial_y \theta $}
\label{eq: 20}
\end{equation}

while, the equation for $\delta S$ (Eq. (2) main text) in homogeneous steady state gives,

\begin{equation}
\centerline{$0=[\alpha_1(\rho_0)+{\alpha_1}^{'}(\rho_0)\delta \rho - \frac{\alpha_2}{2}({S_0}^2+2 S_0 \delta S)]\frac{S_0+\delta S}{2}+..... $}
\label{eq: 21}
\end{equation}

also, $\alpha_1(\rho_0)- \frac{\alpha_2}{2}{S_0}^2=0$. Therefore we have, 
\begin{equation}
\centerline{$({\alpha_1}^{'}(\rho_0)\delta \rho - \alpha_2 S_0 \delta S)(S_0+\delta S)=0$}
\label{eq: 22}
\end{equation}

\begin{equation}
\centerline{$\delta S = \frac{{\alpha_1}^{'}(\rho_0)\delta \rho}{\alpha_2 S_0}$}
\label{eq: 23}
\end{equation}
or,
\begin{equation}
\centerline{$\delta S=\Gamma \delta \rho$}
\label{eq: 24}
\end{equation}

where, $\Gamma=\frac{{\alpha_1}^{'}(\rho_0)}{\alpha_2 S_0}$ and ${\alpha_1}^{'}=\frac{\partial \alpha_1(\rho)}{\partial \rho}\vert_{\rho=\rho_0}$. Hence, Eq. (\ref{eq: 20}) can be re-written as,

\begin{equation}
\centerline{$\partial_t \delta \rho=(\frac{a_0S_0}{2}+D_{\rho}) \partial_{x}^2 \delta \rho +(D_{\rho}-\frac{a_0S_0}{2}) \partial_{y}^2 \delta \rho +\frac{a_0\rho_0}{2} \Gamma(\partial_{x}^2-\partial_{y}^2)\delta \rho+2a_0\rho_0 S_0 \partial_x \partial_y \theta $}
\label{eq: 25}
\end{equation}
or,
\begin{equation}
\centerline{$\partial_t \delta \rho=(\frac{a_0S_0}{2}+D_{\rho}+\Gamma \frac{a_0\rho_0}{2})\partial_{x}^2 \delta \rho + (-\frac{a_0S_0}{2}+D_{\rho}-\Gamma \frac{a_0\rho_0}{2})\partial_{y}^2 \delta \rho+2a_0\rho_0 S_0 \partial_x \partial_y \theta $}
\label{eq: 26}
\end{equation}

or,
\begin{equation}
\centerline{$\partial_t \delta \rho=K_1 \partial_{x}^2 \delta \rho + K_2 \partial_{y}^2 \delta \rho+K_3 \partial_x \partial_y \theta $}
\label{eq: 27}
\end{equation}

Where, $K_1=(\frac{a_0S_0}{2}+D_{\rho}+\Gamma \frac{a_0\rho_0}{2})$, $K_2=(-\frac{a_0S_0}{2}+D_{\rho}-\Gamma \frac{a_0\rho_0}{2})$ and $K_3=2a_0\rho_0 S_0$. 

Now the equation of motion for $\mathcal{Q}_{12}$,

\begin{equation}
\centerline{$\partial_{t}\mathcal{Q}_{12}=[\alpha_{1}(\rho)-\alpha_{2}(\mathcal{Q} : \mathcal{Q})]\mathcal{Q}_{12} +\beta(\nabla_{1}\nabla_{2} - \frac{1}{2}\delta_{12}\nabla^2) \rho + D_{\mathcal{Q}}\nabla^2 \mathcal{Q}_{12}+H_{12} + { \Omega}_{12}$}
\label{eq: 28}
\end{equation}

Here, $\mathcal{Q}_{12}=S_0 \theta$ therefore, in linear order, $[\alpha_{1}(\rho)-\alpha_{2}(\mathcal{Q} : \mathcal{Q})]\mathcal{Q}_{12}$ will not survive. $h_1 h_2={h_0}^2cos \phi sin\phi = {h_0}^2 \Phi({\bf r})$, where $\Phi({\bf r})=cos \phi sin\phi$.

\begin{equation}
\centerline{$\partial_t \theta = \frac{\beta}{\rho_0 S_0}\partial_{x}\partial_{y}\delta \rho + D_{\mathcal{Q}}(\partial_{x}^2+\partial_{y}^2)\theta +\frac{{h_0}^2}{\rho_0 S_0}\Phi +\frac{1}{\rho_0 S_0} {{\Omega}}$}
\label{eq: 29}
\end{equation}

Taking the Fourier transform of equation (\ref{eq: 27}) and (\ref{eq: 29}), where Fourier modes are defined as, $f({\bf{q}},\omega)=\int \int f({\bf{r}},t) e^{i{\bf{q}}\cdot {\bf{r}}+i \omega t}d{\bf{r}}dt$, we get,

\begin{equation}
\centerline{$(K_1 {q_x}^2+K_2 {q_y}^2-i\omega)\delta \rho({\bf{q}},\omega)+K_3 q_x q_y \theta({\bf{q}},\omega)=0$}
\label{eq: 30}
\end{equation}

and,
\begin{equation}
\centerline{$\frac{\beta}{\rho_0 S_0}q_xq_y \delta \rho({\bf{q}},\omega) +[D_{\mathcal{Q}} ({q_x}^2+{q_y}^2)-i\omega]\theta({\bf{q}},\omega)=\frac{{h_0}^2}{\rho S_0}\Phi({\bf{q}}) + \frac{1}{\rho S_0}{{\tilde{\Omega}}}({\bf{q}},\omega)$}
\label{eq: 31}
\end{equation}

Solving equation (\ref{eq: 30}) and (\ref{eq: 31}) will give,

\begin{equation}
\centerline{${\bf{M}} \left[ \begin{array}{c} \delta \rho({\bf{q}},\omega) \\ \theta({\bf{q}},\omega) \end{array} \right]=\frac{1}{\rho_0 S_0}\left[ \begin{array}{c} 0 \\ {h_0}^2 \Phi({\bf{q}})+{{\tilde{\Omega}}}({\bf{q}},\omega) \end{array} \right]$}
\label{eq: 32}
\end{equation}

where,
\begin{equation}
\centerline{${\bf{M}} =\; \left[ \begin{array}{cc} K_1{q_x}^2+K_2{q_y}^2-i\omega & K_3 q_x q_y \\ \frac{\beta}{\rho_0 S_0}q_x q_y & D_{\mathcal{Q}}({q_x}^2+{q_y}^2)-i\omega \end{array} \right]\; $}
\label{eq. 33}
\end{equation}

by solving equation (\ref{eq: 32}) for $q_x=q_y$, we get

\begin{equation}
\centerline{$ \left[ \begin{array}{c} \delta \rho({\bf{q}},\omega) \\ \theta({\bf{q}},\omega) \end{array} \right]=\frac{1}{(D_1q^4+{\omega}^2)-i \omega D_2q^2}\left[ \begin{array}{c} -K_3q^2 \\ 2D_{\rho}q^2-i\omega \end{array} \right]\frac{({h_0}^2 \Phi({\bf{q}})+{{\tilde{\Omega}}}({\bf{q}},\omega))}{\rho_0 S_0}$}
\label{eq: 34}
\end{equation}

where, $D_1=4D_{\rho}D_{\mathcal{Q}}+2a_0\beta$ and $D_2=2(D_{\rho}+D_{\mathcal{Q}})$. Equation (\ref{eq: 34}) gives,

\begin{equation}
\centerline{$\delta \rho({\bf{q}},\omega)=\frac{-K_3 q^2}{(D_1q^4+{\omega}^2)-i \omega D_2q^2}\frac{(h_{0}^{2} \Phi({\bf{q}})+{{\tilde{\Omega}}}({\bf{q}},\omega))}{\rho_0 S_0}$}
\label{eq: 35}
\end{equation}

\begin{equation}
\centerline{$\theta({\bf{q}},\omega)=\frac{2D_{\rho}q^2-i\omega}{(D_1q^4+{\omega}^2)-i \omega D_2q^2}\frac{(h_{0}^{2} \Phi({\bf{q}})+{{\tilde{\Omega}}}({\bf{q}},\omega))}{\rho_0 S_0}$}
\label{eq: 36}
\end{equation}

Now, we first calculate the two point orientation correlation functions,  

\begin{equation}
\centerline{$\langle\theta({\bf{q}},\omega)\theta({\bf{-q}},-\omega) \rangle = \frac{{D_{\rho}}^2q^4+\omega^2}{(D_1q^4+{\omega}^2)^2+\omega^2 {D_2}^2q^4}\frac{[h_{0}^{4}\langle \Phi({\bf{q}})\Phi({\bf{-q}}) \rangle + \langle {{\tilde{\Omega}}}({\bf{q}},\omega){{\tilde{\Omega}}}({\bf{-q}},-\omega) \rangle]}{\rho_0 S_0}$}
\label{eq: 37}
\end{equation}

here, $\langle \Phi({\bf{q}})\Phi({\bf{-q}}) \rangle =\delta({\bf{q+q}})$ and $\langle {{\tilde{\Omega}}}({\bf{q}},\omega){{\tilde{\Omega}}}({\bf{-q}},-\omega) \rangle = \bigtriangleup_0 \delta({\bf{q+q}})\delta(\omega +\omega)$. Using this we get,

\begin{equation}
\centerline{$S_{q}(\theta)=\mathcal{C}(D_{\rho},D_{\mathcal{Q}})\frac{h_{0}^{4}}{q^4}+\mathcal{B}(D_{\rho},D_{\mathcal{Q}})\frac{1}{q^2}$}
\label{eq: 38}
\end{equation}

Where, $\mathcal{C}(D_{\rho},D_{\mathcal{Q}})=\frac{4{D_{\rho}}^2}{\rho_0 S_0(4D_{\rho}D_{\mathcal{Q}}+2a_0\beta)^2}$ and $\mathcal{B}(D_{\rho},D_{\mathcal{Q}})=\frac{\pi \bigtriangleup_0 }{2\rho_0 S_0}\frac{1}{c\sqrt{2(4b^2+c^2)}}[\frac{(2{D_{\rho}}^2+c(\sqrt{4b^2+c^2})-2b^2)}{\sqrt{c(c-\sqrt{4b^2+c^2})+2b^2}}+\frac{(-2{D_{\rho}}^2+c(\sqrt{4b^2+c^2})+2b^2)}{\sqrt{c(c+\sqrt{4b^2+c^2})+2b^2}}]$, where $ b=\sqrt{2(2D_{\rho}D_{\mathcal{Q}}+a_0\beta)} \ and \ c=\sqrt{2(D_{\rho}+D_{\mathcal{Q}})}$.
\end{widetext}
Hence, the two point angle correlation function can be written as,

\begin{equation}
\centerline{$S_q(\theta) \simeq \frac{\mathcal{B}}{q^2}+\frac{\mathcal{C} h_{0}^{4}}{ q^4}$}
\label{eq: 39}
\end{equation}

Here, the coefficients $\mathcal{C}$ and $\mathcal{B}$ depends on system parameters.
To get the two point correlation function for nematic orde parameter $C_{\mathcal{Q}}(x) \simeq exp(-G_{\theta}(x))$ \cite{chaikinlubenskey}, where $G_{\theta}(x)$ is the inverse Fourier transform of $S_q(\theta) $ Eq. (\ref{eq: 39}). Also, $G(x)=\mathcal{B} f(x)+ \mathcal{C} h_{0}^{4} g(x)$, where, 

\begin{equation}
\centerline{$f(x)= \int_{2\pi/L}^{2\pi/a}\frac{d^2q}{4\pi^2}\frac{1-e^{i{\bf{q\cdot x}}}}{q^2} \simeq ln(\Lambda \vert x \vert)$}
\end{equation}

and,

\begin{equation}
\centerline{$ g(x)=\int_{2\pi/a}^{2\pi/L} \frac{dq}{q^3}[\frac{1}{2}\int_{0}^{2\pi}d\theta (1-e^{iq\vert x \vert cos\theta})] $}
\end{equation}
or,
\begin{equation}
\centerline{$ g(x)=\int_{2\pi/a}^{2\pi/L} \frac{dq}{q}(1-J_0(q\vert x \vert) $}
\end{equation}
here, $J_n$ is the $n^{th}$ order Bessel's function \cite{arfkenandweber}.

\begin{equation}
\centerline{$ g(x)=\vert x \vert^2 \int_{0}^{1} \frac{du(1-J_0(u))}{u^3} + \vert x \vert^2 \int_{1}^{\Lambda \vert x \vert} \frac{du}{u^3}-\vert x \vert^2 \int_{1}^{\Lambda \vert x \vert}  \frac{du J_0(u))}{u^3} $}
\end{equation}

\begin{equation}
\centerline{$ g(x)=\vert x \vert^2 A + \vert x \vert^2 [-\frac{1}{2}(1-\frac{1}{\Lambda^2 \vert x \vert^2} )]-\vert x \vert^2 \int_{1}^{2\pi/a \vert x \vert}  \frac{du J_0(u))}{u^3} $}
\end{equation}

\begin{equation}
\centerline{$ g(x)=\vert x \vert^2 (A -\frac{1}{2}-\vert x \vert^2 \int_{1}^{2\pi/a \vert x \vert}  \frac{du J_0(u))}{u^3}) $}
\end{equation}

\begin{equation}
\centerline{$g(x)=\frac{a^2}{2\pi^2}+\vert x \vert^2 (A -\frac{1}{2}-A') $}
\end{equation}

here, $A=\int_{2\pi/L}^{2\pi/a}\frac{1-J_0(u)}{u^3}du \simeq 1.2$, $A'=\int_{1}^{2\pi/a \vert x \vert}\frac{J_0(u)}{u^3}du \simeq \int_{0}^{\infty}\frac{J_0(u)}{u^3}du \simeq 0.27$. Here, $a=1$ is the lattice spacing. 

\begin{equation}
\centerline{$ g(x)=\vert x \vert^2 \times \mathcal{O}(0.01) $}
\end{equation}

Hence, the orientation correlation function is given by,

\begin{equation}
\centerline{$ C(x)\simeq \frac{1}{\vert x \vert^{\mathcal{B}}} e^{-\vert x \vert^2 \times \mathcal{O}(0.01) \times \mathcal{C} h_{0}^{4}} $}
\end{equation}

when measured on the scale of system size $N=K^2$, we get, 

\begin{equation}
\centerline{$C_{\mathcal{Q}}(N) \simeq \frac{1}{N^\mathcal{B'}} e^{-\mathcal{C'}h_{0}^{4} N}$}
\label{eq: 42}
\end{equation}

Here, $\mathcal{B}'=1.17 \times 10^{-4}$ and $\mathcal{C'}=3.9 \times 10^{-3}$.\\

Similarly, structure factor for density can be calculated using Eq. \ref{eq: 35} and given by,
 
\begin{equation}
\centerline{$S_{\rho}({\bf q})= \gamma_1 \frac{h_0^4}{q^4}+\gamma_2 \frac{\bigtriangleup_0}{q^2}$}
\label{eq: 42}
\end{equation} 
where, $\gamma_1=0.5$ and $\gamma_2=0.4$ are constants and depends only on system parameters.

\section{Snapshots for $\Delta \theta$ and NOP}
\label{appendix}
\subsection{Snapshots for $\Delta \theta$}
\begin{figure*}
\includegraphics[height=3cm, width=12cm]{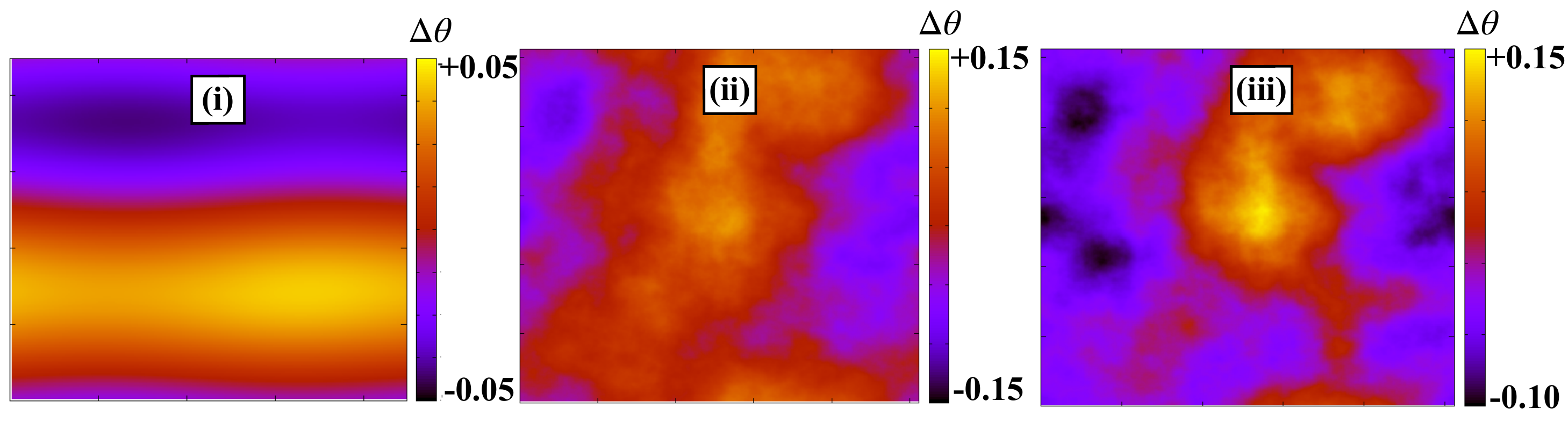}
\caption{(Color online) Snapshots of $\Delta \theta$ for $h_0=0.0, a_0=0.2$ (i), $h_0=0.05, a_0=0.2$ (ii) and $h_0=0.05, a_0=0.3$ (iii).}

\label{fig: 8}
\end{figure*}

The snapshots corresponding to Fig. \ref{fig: 2} (c) is shown in Fig. \ref{fig: 8}. Here we can see that for a non-zero disorder in the system; distinct domains can be seen as 
the fluctuation in angular orientation $\Delta \theta$ represented by color bar varies significantly throughout the space whereas, for the clean system, the whole space is identical in terms of $\Delta \theta$.
Also for larger activity $a_0=0.3$, the magnitude of $\Delta \theta$ fluctuations decreases, which confirms the stronger intra-domain ordering as found in $P(\Delta \theta)$.\\
\subsection{Fixed $a_0$ and varying  $h_0$}

\begin{figure*}
\includegraphics[height=6cm, width=12cm]{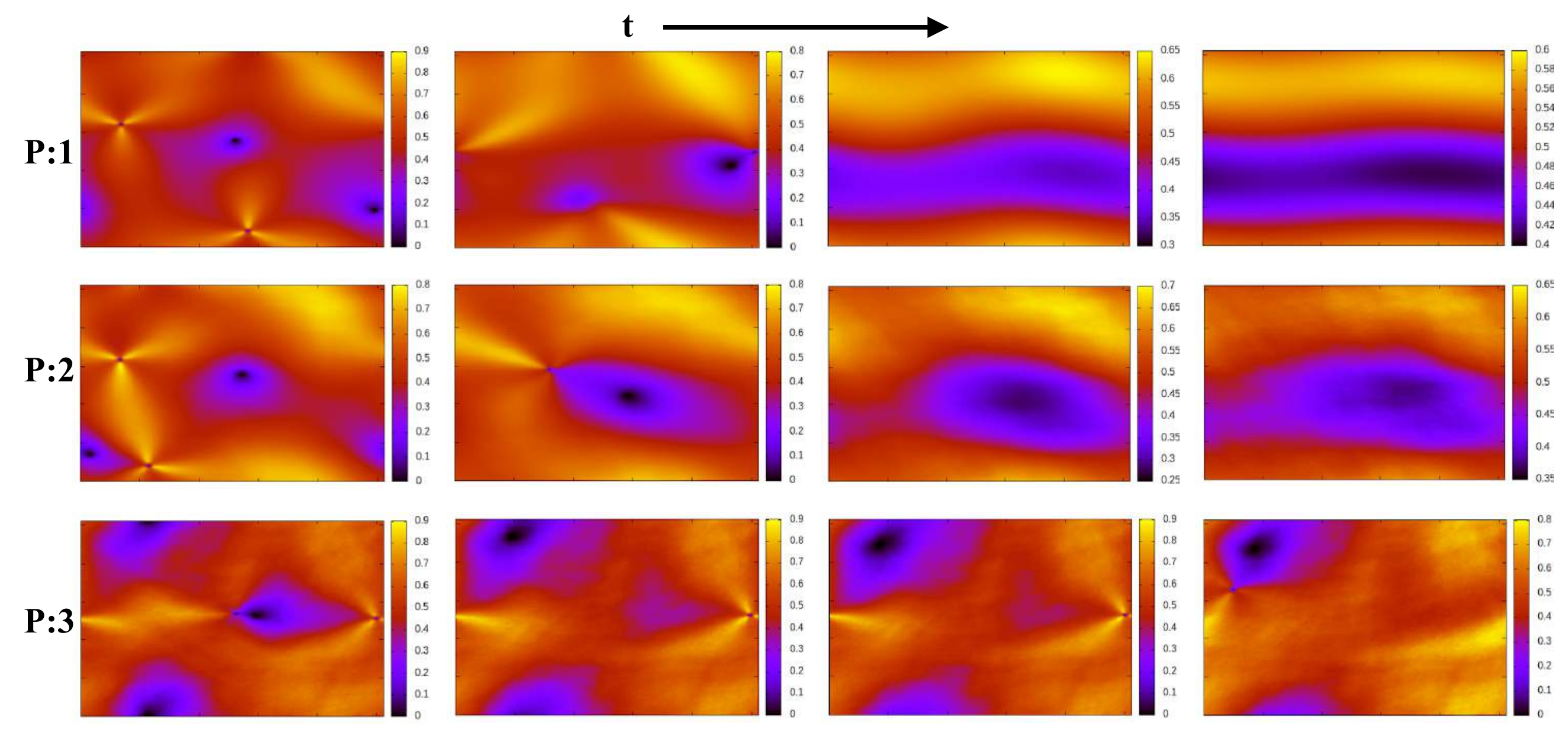}
	\caption{(Color online) Snapshots of local NOP, $\mathcal{Q}$ for $a_0=0.2$ and for different disorder strength in the system. From top to bottom  panel (P:1 to P:3) $h_0 = 0.0$, $0.05$ and $0.1$ respectively. Snapshots are generated at equal interval i.e. $t=150000, \ 300000, \ 450000, \ and \ 600000$ from left to right respectively. Notice the numbers on the color bars.}
\label{fig: 9}
\end{figure*}

In fig \ref{fig: 9}, we show the snapshots for local NOP, $\mathcal{Q}$ at 
different simulation time for $a_0=0.2$ and different strengths of 
disorder in the system. 
We also included the multimedia files in the SM (see the supplementary materiel for the animations) for the same. We observe that  as we increase the disorder in the system, 
dynamics of defect is slows down. Also for high disorder, defects are pinned, 
which is responsible for formation of 
multiple smaller domains as shown in Fig. \ref{fig: 2}(c) and \ref{fig: 8}(i-iii). \\

\subsection{Fixed disorder $h_0$ and varying activity $a_0$}
\begin{figure*}
\includegraphics[height=3cm, width=8cm]{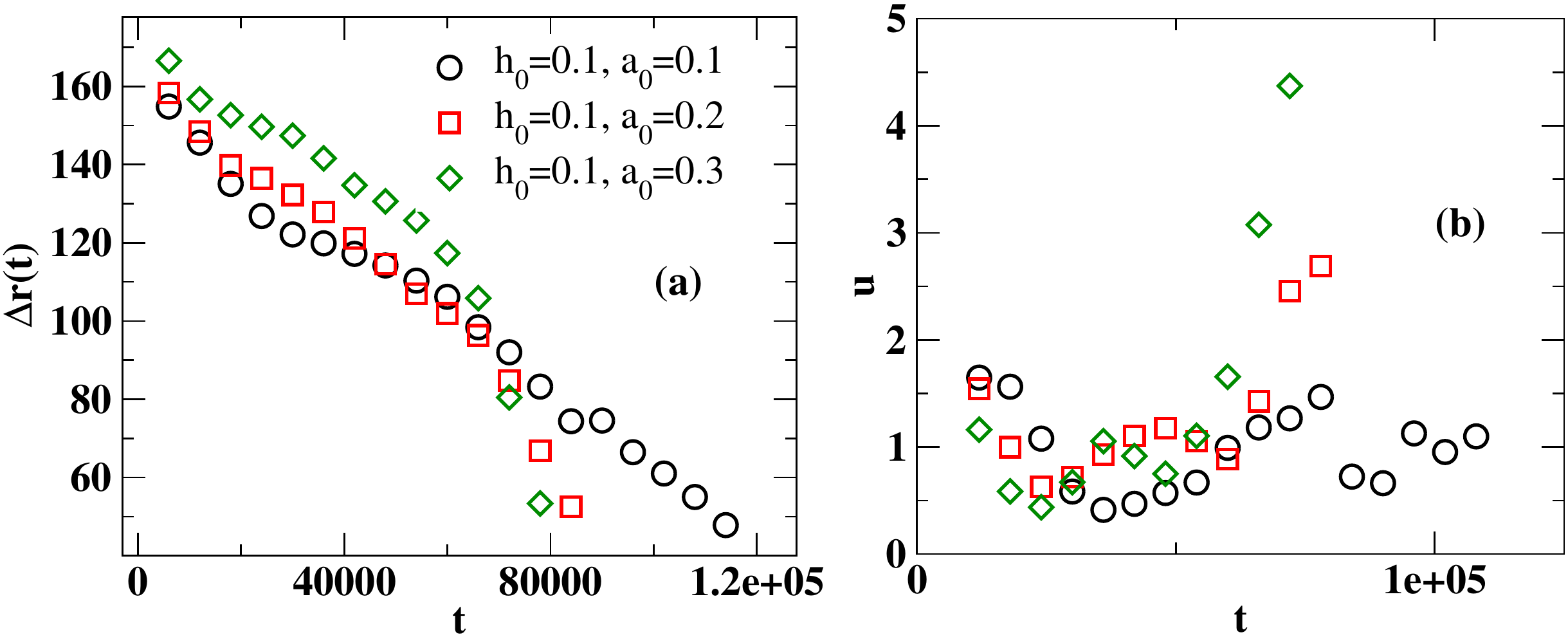}
\caption{(Color online) (a) Relative separation $\Delta (r)$ between $+1/2$ and $-1/2$ defects pair  vs. time plot and (b) relative  speed, $u=|\frac{d}{dt}\Delta r(t)|\times 10^{-3}$, of $\pm 1/2$ defects, for different value of $a_0$ and fixed $h_0=0.1$.  $t$ is the simulation time.}

\label{fig: 11}
\end{figure*}

Further, we change activity  $a_0$ in equation (1)  (main text), 
and plot the snapshots of local NOP, $\mathcal{Q}$ for fixed $h_0=0.05$ in fig \ref{fig: 10}
(see the supplementary materiel for animations). We find that, for a fixed  $h_0$ ($=0.05$ in this case), as we increase  $a_0$, 
annihilation of defects happens faster than that of for the the smaller $a_0$. We also  plot the relative separation $\Delta r(t)$  between a $+1/2$ and 
$-1/2$ defects {\em vs.} $t$ in  fig. \ref{fig: 11}(a) for three
different $a_0=0.1$, $0.2$ and $0.3$. Also the relative speed, which is defined as, $u=|\frac{d}{dt}\Delta r(t)|\times 10^{-3}$, is plotted in fig \ref{fig: 11}(b).\\

\begin{figure*}
\includegraphics[height=6cm, width=12cm]{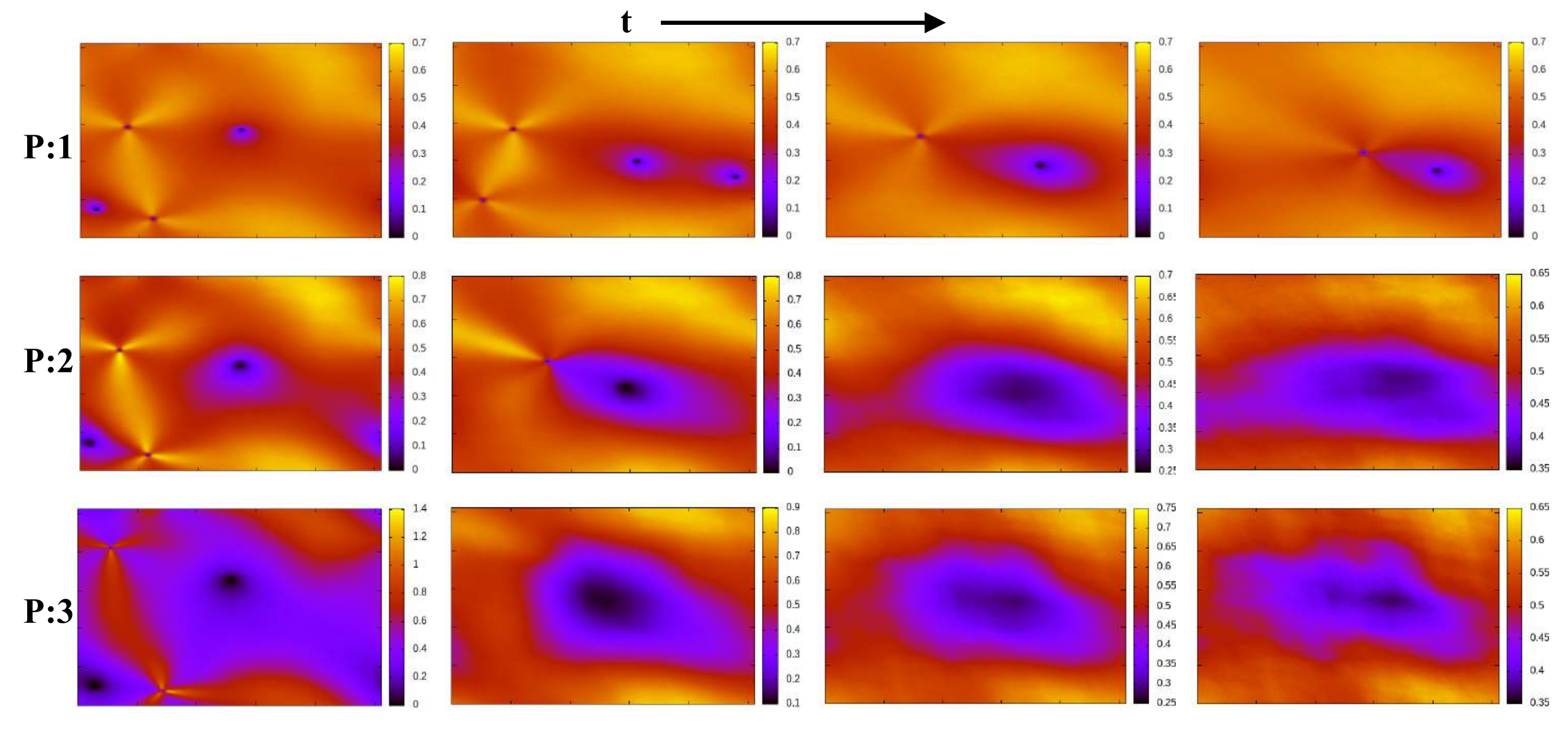}
	\caption{(Color online) Snapshots of local NOP, $\mathcal{Q}$ for $h_0=0.05$ and for different activity in the system. From top to bottom  panel (P:1 to P:3) $a_0 = 0.1$, $0.2$ and $0.3$ respectively. Snapshots are generated at equal interval i.e. $t=150000, \ 300000, \ 450000, \ and \ 600000$ from left to right respectively. Notice the numbers on the color bars.}
\label{fig: 10}
\end{figure*}

\end{document}